\documentclass[12pt]{iopart}
\usepackage{times}
\usepackage{t1enc}
\usepackage[english]{babel}
\usepackage{graphicx}
\usepackage{relsize}
\usepackage{iopams}
\usepackage{mathptmx}
\usepackage{xcolor}
\newcommand{\iu}{{i\mkern1mu}}
\eqnobysec

\begin{document}
\title{Picosecond-level corrections to the Shapiro time delay in a Schwarzschild spacetime}

\author{Oscar del Barco$^{*}$}

\address{Departamento de F\'{i}sica, Instituto
Universitario de Investigaci\'{o}n en \'{O}ptica y
Nanof\'{i}sica, Universidad de Murcia, Campus de
Espinardo, E-30100, Murcia, Spain}
\address{$^*$ Author to whom any correspondence should be addressed.}
\ead{obn@um.es}

\begin{abstract}
We present an exact analytical equation for the Shapiro time
delay (STD) due to a spherical non-rotating body.
As a result, accurate values of the STD in comparison
with first and second-order expressions for Schwarzschild
spacetime (1Sch and 2Sch) and first-order post-Newtonian
formalism (1PN) are achieved. Accordingly, the lowest STD
discrepancies between our exact equation and these approximations
lie within the picosecond and sub-picosecond
level for light beams affected by the Sun's gravity.
Our results might be useful for time delay measurements
in the solar system or extragalactic binary
pulsar systems, where a high accuracy level is required.
\end{abstract}

\maketitle

\section{Introduction}

The time delay of a light beam in the gravitational field
of a massive body constitutes one of the traditional tests
of general relativity (GR), together with the perihelion
precession of Mercury, gravitational deflection of light
and gravitational redshift. Although this effect
is a natural consequence of the equivalence principle,
the gravitational time delay was not predicted until
1964 by Irwin Shapiro \cite{SH64} in the context
of timing radar pulses reflected from Venus and Mars at
superior conjunction. In addition to checking the GR in
the solar system \cite{SH68,SH71}, the STD has also been
applied to measure the speed of gravitational interaction
\cite{KO01,KO03,WI14}, verify modiﬁed
gravity theories \cite{AS08,ED21,DY22}
and investigate astrophysical objects such as pulsars
\cite{HA19,CR20,BE22}. This relativistic effect can also be
analyzed in the context of the material medium approach (MMA),
where the gravitational field is represented as an optical medium
with an effective refractive index, providing a satisfactory
explanation for the deflection of electromagnetic waves by a gravitational
field \cite{BA58,PL60,DE71,NA95,EV96a,EV96b,SE10,RO15,RO17,FE19,FE20,BA24,RU25,RO25}.
Moreover, the STD was also computed numerically using the MMA
approach \cite{FE19,BA24}, where results consistent
with GR calculations were achieved.

On the other hand, an exact analytical equation for the
Shapiro time delay in Schwarzschild spacetime
may be obtained in terms of incomplete elliptic
integrals of different types \cite{BA24}.
In this sense, analytical expressions for the
propagation time delay in the timing of
pulsars orbiting supermassive black holes have
been reported \cite{HA19,BE22},
where elliptic integrals of the first, second, and
third types are involved. In this article,
we derive an exact analytical equation for the STD due to
a static massive object, revealing the equivalence
between the MMA approach and GR formalism when
studying light propagation in a gravitational
field. Consequently, the lowest differences between
our analytical formula and previous first and second-order approximate
expressions lie within the picosecond and sub-picosecond level.
We believe that our STD equation might be a helpful tool to
perform ultraprecise time delay calculations
for both solar system and extragalactic scenarios,
provided that picosecond \cite{WI14} and even
sub-picosecond accuracy levels \cite{ZS22,ZS24}
are essential for current and potential STD measurements.
It is also expected that future gravitational
detectors will enhance this accuracy level
\cite{BA10,SU20}.

The paper is organized as follows. In Section 2,
we deduce our exact analytical equation for the Shapiro time delay
in a Schwarzschild spacetime in terms of elliptical functions.
Furthermore, the gravitational deflection angle of light (GDA)
is also revisited and two accurate formulas (each one expressed
in an adequate coordinate system, i.e., Schwarschild or isotropic
coordinates) are described. In Section 3, we present our
fundamental numerical results for both STD and GDA in the solar system,
highlighting the suitability of our exact expressions
in ultraprecise time delay calculations. Finally, we summarize our main
results and conclusions in Section 4, and include two Appendices with
detailed calculations related to the exact and second-order STD equations.

\section{Theoretical basis}

Consider the astrophysical scenario described
in Figure \ref{fig1}. A light beam is emitted from
the Earth (E), passes near the Sun (S) at the closest
distance or perihelion $r_0$ and is reflected
back from planet P. Accordingly, the coordinate time $t$ for a light beam
to propagate from $r_0$ to a specific Schwarzschild
coordinate $r$ obeys the following ordinary differential
equation in the equatorial plane \cite{WE72,WA84}
\begin{equation}\label{DEGR}
c\frac{\mathrm{d}t}{\mathrm{d}r} = \left(1-\frac{r_{\rm s}}{r}\right)^{-1}
\left[1-\left(1-\frac{r_{\rm s}}{r}\right) \frac{b^2}{r^2}\right]^{-1/2},
\end{equation}
where $b$ stands for the impact parameter of the light pulse
\begin{equation}\label{bvsr0}
b^2= r_0^2 \left(1-\frac{r_{\rm s}}{r_0}\right)^{-1}.
\end{equation}

Therefore, equation (\ref{DEGR}) can be rewritten as
\begin{equation}\label{DEGRdef}
c\frac{\mathrm{d}t}{\mathrm{d}r} = \left(1-\frac{r_{\rm s}}{r}\right)^{-1}
\left[1-\left(\frac{r_0}{r}\right)^{3} \left(\frac{r-r_{\rm s}}
{r_0-r_{\rm s}}\right)\right]^{-1/2},
\end{equation}
which is equivalent to equation (24)
in \cite{BA24} deduced via the MMA method.
Consequently, this optics-based
formalism reproduces exactly the time delay in a
gravitational field due to a static massive object.
Moreover, analytical solutions for the STD in terms
of incomplete elliptic integrals of different
types are found, as briefly discussed.

\subsection{Exact analytical equation for the Shapiro time delay}

\begin{figure}
\begin{center}
\includegraphics[width=.77\textwidth]{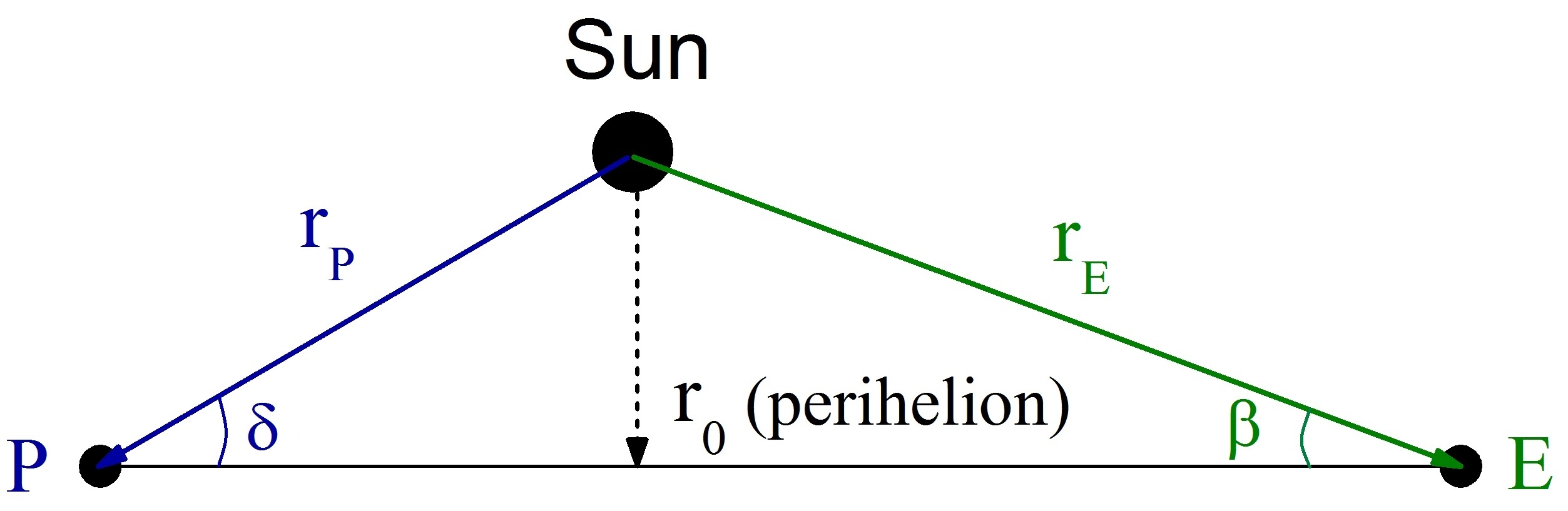}
\caption{Our astrophysical scenario for the
Shapiro time delay analysis in a Schwarzschild
spacetime. A light beam is emitted from the Earth (E),
passes near the Sun (S) at the closest distance or
perihelion $r_0$ and is reflected back from planet P.}\label{fig1}
\end{center}
\end{figure}

As shown in Figure \ref{fig1}, the STD
for an Earth-planet round trip can be
easily expressed as
\begin{equation}\label{Shaptimed}
\Delta t = 2\left[(\Delta t)_{r_0 \to r_{\rm P}}+
(\Delta t)_{r_0 \to r_{\rm E}}-\frac{1}{c}
\left(\sqrt{r_{\rm P}^2-r_0^2}+
\sqrt{r_{\rm E}^2-r_0^2}\right)\right],
\end{equation}
where $(\Delta t)_{r_0 \to r_{\rm P}}$ and
$(\Delta t)_{r_0 \to r_{\rm E}}$
represent the propagation times of a light
beam from its perihelion $r_0$ to planet P
or Earth, respectively. The general form
of $(\Delta t)_{r_0 \to r}$ can be analytically
evaluated from the direct integration of
equation (\ref{DEGRdef}), and corresponds
to equation (\ref{coordtimer0rdeff}) in Appendix A.

Let us recall that we are considering the coordinate
time delay $\Delta t$, which is not the time an
Earth-based clock would measure. In all cases, we
must account for the gravitational influence
of the Sun as well as Earth’s translational motion.
Therefore, the proper time delay $\Delta\tau$ elapsed between
the emission and arrival of the light beam on Earth
is finally \cite{MA24}
\begin{eqnarray}\label{Shaptimedproper}
\Delta\tau = 2\left[(\Delta t)_{r_0 \to r_{\rm P}}+
(\Delta t)_{r_0 \to r_{\rm E}}-\frac{1}{c}
\left(\sqrt{r_{\rm P}^2-r_0^2}+
\sqrt{r_{\rm E}^2-r_0^2}\right)
\left(1+\frac{3r_{\rm s}}{4r_{\rm E}}\right)\right].
\nonumber\\
\end{eqnarray}
Throughout this article, we restrict ourselves to
coordinate time delay calculations $\Delta t$ rather
than a proper time delay analysis. This choice
does not affect our STD comparisons with
previous approximate expressions.

In this respect, the first-order STD equation in
Schwarzschild coordinates can be found in many standard GR textbooks \cite{WE72,WA84}
\begin{eqnarray}\label{Shap1stGR}
\Delta t^{(\rm 1Sch)}&=&\frac{2r_{\rm s}}{c}
\left[\log\left(\frac{r_{\rm P}+\sqrt{r_{\rm P}^{2}-r_0^2}}{r_0}\right)+
\log\left(\frac{r_{\rm E}+\sqrt{r_{\rm E}^{2}-r_0^2}}{r_0}\right)\right]
            \nonumber\\
&+& \frac{r_{\rm s}}{c} \left(\sqrt{\frac{r_{\rm P}-r_0}{r_{\rm P}+r_0}}
+\sqrt{\frac{r_{\rm E}-r_0}{r_{\rm E}+r_0}}\right),
\end{eqnarray}
while a second-order expression for the Shapiro time delay
can be easily derived after an adequate Taylor expansion up to $r_{\rm s}^{2}$
of the right-hand side (RHS) of equation (\ref{DEGRdef})
(please, see the details in Appendix B, in particular, equation (\ref{Delta2Schfin}))
\begin{eqnarray}\label{Shap2ndGR}
\Delta t^{(\rm 2Sch)}&=&\Delta t^{(\rm 1Sch)}+\frac{r_{\rm s}^{2}}{4cr_0}
\left[15\arctan\left(\frac{\sqrt{r_{\rm P}^{2}-r_0^{2}}}{r_0}\right)
-\left(\frac{4r_{\rm P}+5r_0}{r_{\rm P}+r_0}\right)
\sqrt{\frac{r_{\rm P}-r_0}{r_{\rm P}+r_0}}\right.
\nonumber\\
&+& \left.15\arctan\left(\frac{\sqrt{r_{\rm E}^{2}-r_0^{2}}}{r_0}\right)
-\left(\frac{4r_{\rm E}+5r_0}{r_{\rm E}+r_0}\right)
\sqrt{\frac{r_{\rm E}-r_0}{r_{\rm E}+r_0}}\right].
\end{eqnarray}
Considering that our exact equation (\ref{Shaptimed}) is also
expressed in the same coordinate system, a direct
comparison between previous formulas is straightforward.

Additionally, given that the isotropic coordinate system
is most appropriate for astrophysical observational
purposes, it is advisable to represent our analytical
equation (\ref{Shaptimed}) in terms of isotropic
coordinates. Taking into account the well-known
transformation relations \cite{MI73},
\begin{equation}\label{transr0}
r_0=r_{0 \rm I} \left(1+\frac{r_{\rm s}}{4r_{0 \rm I}}\right)^{2},
\end{equation}
\begin{equation}\label{transrP}
r_{\rm P}=r_{\rm PI} \left(1+\frac{r_{\rm s}}{4r_{\rm PI}}\right)^{2},
\end{equation}
\begin{equation}\label{transrE}
r_{\rm E}=r_{\rm EI} \left(1+\frac{r_{\rm s}}{4r_{\rm EI}}\right)^{2},
\end{equation}
where $r_{0 \rm I}$, $r_{\rm PI}$ and $r_{\rm EI}$ represent the heliocentric
isotropic coordinates of the perihelion, planet P and the Earth, respectively,
we can now compute exactly the Shapiro time delay via our
analytical formula in this coordinate system. To this end,
we introduce equations (\ref{transr0}), (\ref{transrP}) and (\ref{transrE})
into the STD equations (\ref{Shaptimed}) and (\ref{coordtimer0rdeff})
in Appendix A.

Although an explicit form of equation (\ref{Shaptimed})
in isotropic coordinates is not included in this article
(owing to its notable analytical complexity),
all subsequent STD calculations in this coordinate system
will be performed upon the aforementioned coordinate
transformation. In addition, these considerations are required
to properly compare our exact STD results with previous
first-order approximate expressions based on the
post-Newtonian formalism, which are commonly expressed
in isotropic or harmonic coordinates. Thus, the 1PN formula
for the Shapiro time delay according to Misner \cite{MI73}
or Poisson \cite{PO14} reads
\begin{equation}\label{Shap1PNisoMis}
\Delta t_{\rm Mis}^{(\rm 1PN)}= \frac{2r_{\rm s}}{c}
\left[\log\left(\frac{r_{\rm PI}+\sqrt{r_{\rm PI}^{2}-r_0^2}}{r_0}\right)+
\log\left(\frac{r_{\rm EI}+\sqrt{r_{\rm EI}^{2}-r_0^2}}{r_0}\right)\right],
\end{equation}
where, as mentioned, the positions of planet P and the
Earth are given in heliocentric isotropic coordinates.

Alternatively, the 1PN equation for the STD can be
expressed as \cite{MA24}
\begin{eqnarray}\label{Shap1PNisoMal}
\Delta t_{\rm Mal}^{(\rm 1PN)}&=&\frac{2r_{\rm s}}{c}
\left[\log\left(\frac{r_{\rm PI}+\sqrt{r_{\rm PI}^{2}-r_0^2}}{r_0}\right)+
\log\left(\frac{r_{\rm EI}+\sqrt{r_{\rm EI}^{2}-r_0^2}}{r_0}\right)\right]
            \nonumber\\
&+& \frac{2r_{\rm s}}{c} \left(\sqrt{\frac{r_{\rm PI}-r_0}{r_{\rm PI}+r_0}}
+\sqrt{\frac{r_{\rm EI}-r_0}{r_{\rm EI}+r_0}}\right),
\end{eqnarray}
which differs from Misner's equation (\ref{Shap1PNisoMis})
by an additional term. It should be emphasized that
the apparent discordance between both 1PN results has an
inherent historical and numerical background, due to the well-known
computational differences between the logarithmic term in
equations (\ref{Shap1PNisoMis}) and (\ref{Shap1PNisoMal}) and
the square root term in Malta's 1PN expression. More precisely,
while these two terms are numerically comparable in magnitude
for $r_0/r \simeq 1$, they drastically differ when $r_0/r \ll 1$.

In the next section, a detailed analysis of our
analytical STD equations and the previous first and
second-order approximate formulas,
considering a suitable coordinate system, is carried out.

\subsection{Gravitational deflection angle of light}

For the sake of completeness, we now review
the gravitational deflection angle of light by a
static massive object. Accordingly,
the exact equation for GDA in Schwarzschild
coordinates $\Delta\alpha$ considering the actual
distances from source and observer
to the gravitational mass is given by
(cf. equation (17) in \cite{BA24})
\begin{eqnarray}\label{GDGOSch}
\Delta\alpha&=&2 \sqrt{\frac{r_0}{Q}} \left[2
\ F\left(\frac{\pi}{2},k\right) - F(z(r_{\rm P}),k)-
F(z(r_{\rm E}),k)\right]
            \nonumber\\
&-& \left[\arccos\left(\frac{r_0}{r_{\rm P}}\right)+
\arccos\left(\frac{r_0}{r_{\rm E}}\right)\right],
\end{eqnarray}
where $Q^2=(r_0-r_{\rm s})(r_0+3r_{\rm s})$
and $F(z,k)$ is the Legendre elliptic integral of
the first kind, with the following relations for
the Jacobi amplitude $z(r)$
\begin{equation}\label{Fzsch}
\sin^2 z(r) = \frac{2r_0 r_{\rm s}+r(r_{\rm s}-r_0+ Q)}
{r(3r_{\rm s}-r_0+Q)},
\end{equation}
and the elliptic modulus $k$
\begin{equation}\label{Fm}
k=\frac{3r_{\rm s}-r_0+Q}{2Q}.
\end{equation}
Moreover, the first-order GR equation
for the gravitational deflection angle of
light in Schwarzschild coordinates is
given by \cite{WE72}
\begin{equation}\label{GDSch1st}
\Delta\alpha^{(\rm 1Sch)}=\frac{r_{\rm s}}{2r_0}
\left(\cos\beta + \cos\delta +
\sqrt{\frac{r_{\rm E}-r_0}{r_{\rm E}+r_0}}+
\sqrt{\frac{r_{\rm P}-r_0}{r_{\rm P}+r_0}}\right),
\end{equation}
where $\sin\beta=r_0/r_{\rm E}$ and $\sin\delta=r_0/r_{\rm P}$
(please, see again Figure \ref{fig1}).

We can also rewrite our exact formula for the GDA
in isotropic coordinates by substituting
equations (\ref{transr0}), (\ref{transrP}) and (\ref{transrE}) into
equations (\ref{GDGOSch}) and (\ref{Fzsch}). After
some elementary algebra, we arrive at the following result
\begin{eqnarray}\label{GDGOiso}
\Delta\alpha&=& 2 \left(1+\frac{r_{\rm s}}
{4r_{0 \rm I}}\right) \sqrt{\frac{r_{0 \rm I}}{Q_{\rm I}}} \ \left[2
\ F\left(\frac{\pi}{2},k_{\rm I}\right) - F(z(r_{\rm PI}),k_{\rm I})-
F(z(r_{\rm EI}),k_{\rm I})\right]
            \nonumber\\
&-& \left[\arccos\left(\frac{r_{0 \rm I}}{r_{\rm PI}}
\left(\frac{1+\frac{r_{\rm s}}{4r_{0 \rm I}}}{1+\frac{r_{\rm s}}{4r_{\rm PI}}}\right)^{2}\right)+
\arccos\left(\frac{r_{0 \rm I}}{r_{\rm EI}}
\left(\frac{1+\frac{r_{\rm s}}{4r_{0 \rm I}}}{1+\frac{r_{\rm s}}{4r_{\rm EI}}}\right)^{2}\right)\right],
            \nonumber\\
\end{eqnarray}
where now
\begin{equation}\label{Qiso}
Q_{\rm I}^{2}=r_{0 \rm I}^{2} \left(1+\frac{r_{\rm s}}
{4r_{0 \rm I}}\right)^{4}-3r_{\rm s}^{2}+2r_{\rm s}r_{0 \rm I}
\left(1+\frac{r_{\rm s}} {4r_{0 \rm I}}\right)^{2},
\end{equation}
\begin{equation}\label{Fmiso}
k_{\rm I}=\frac{1}{2}+\frac{3r_{\rm s}-r_{0 \rm I} \left(1+\frac{r_{\rm s}}
{4r_{0 \rm I}}\right)^{2}} {2Q_{\rm I}},
\end{equation}
and
\begin{eqnarray}\label{Fziso}
\sin^2 z(r_{\rm I})&=&1-\frac{2r_{\rm s}}{Q_{\rm I}+3r_{\rm s}-
r_{0 \rm I} \left(1+\frac{r_{\rm s}}{4r_{0 \rm I}}\right)^{2}}
\nonumber\\
&+& \frac{32 r_{0 \rm I} r_{\rm s} r_{\rm I} \left(1+\frac{r_{\rm s}}{4r_{0 \rm I}}\right)^{2}
}{(r_{\rm s}+4r_{\rm I})^2\left(Q_{\rm I}+3r_{\rm s}-
r_{0 \rm I} \left(1+\frac{r_{\rm s}}{4r_{0 \rm I}}\right)^{2}\right)}.
\end{eqnarray}
Additionally, the first-order equation for the GDA
under the post-Newtonian formalism in isotropic
coordinates reads \cite{LI22a,LI22b}
\begin{equation}\label{GD1PN}
\Delta\alpha^{\rm{(1PN)}}= (1+\gamma) \frac{G M_{\odot}}{r_0 c^2}
\left(\cos\beta + \cos\delta\right),
\end{equation}
where $M_{\odot}$ is the solar mass and
$\gamma$ stands for the dimensionless
post-Newtonian parameter used to
characterize the contribution of the spacetime
curvature to the gravitational deflection.
In this regard, we assume that $\gamma=1$
(as theoretically established in GR) since
this choice does not influence substantially
the results of the GDA.

As in the case of the Shapiro time delay,
a proper comparison between different GDA
expressions with the same coordinate system
is briefly performed.

\section{Results}

In this section, we present some analytical
results concerning the Shapiro time delay and
gravitational deflection angle of light,
as described in the astrophysical scenario
of Figure \ref{fig1}. Consequently,
non-negligible discrepancies between our
exact equations and first and second-order
approximate formulas are reported.

\begin{figure}
\begin{center}
\includegraphics[width=.90\textwidth]{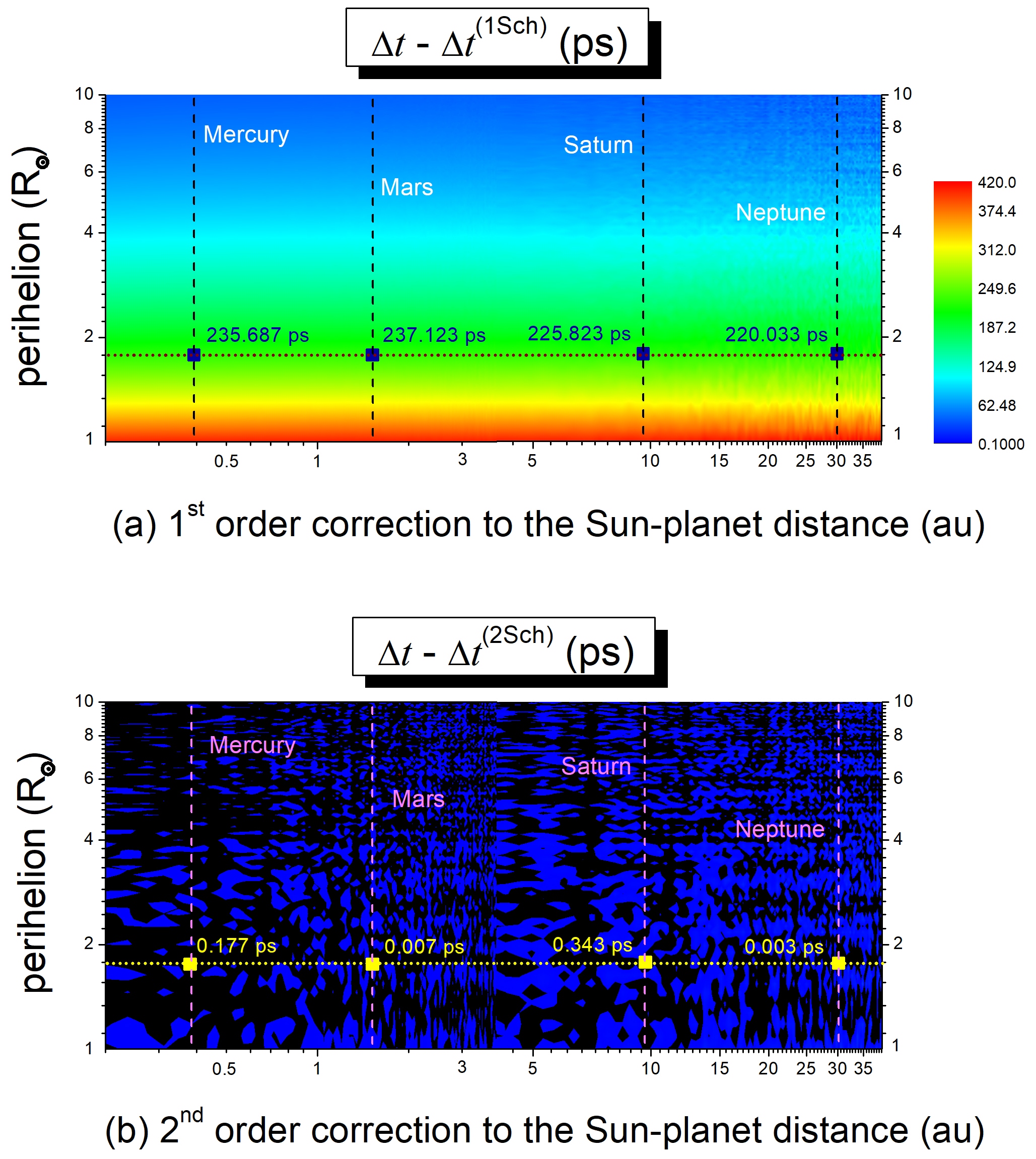}
\caption{(Top panel) Absolute difference between
our exact equation (\ref{Shaptimed}) and the
first-order STD in Schwarzschild coordinates, equation (\ref{Shap1stGR}),
in the case of solar system planets and different perihelion distances.
(Bottom panel) Absolute difference between our analytical
expression (\ref{Shaptimed}) and the second-order STD,
equation (\ref{Shap2ndGR}), for the same astrophysical
situation as in Figure \ref{fig2}(a). It can be observed that the discrepancies
reduce to the sub-picosecond level in the latter case for different
Sun-planet distances and perihelions.}\label{fig2}
\end{center}
\end{figure}

First, Figure \ref{fig2}(a)
represents the absolute difference between
our exact equation (\ref{Shaptimed})
and the first-order STD in Schwarzschild
coordinates, equation (\ref{Shap1stGR}),
in the case of solar system planets
and different perihelion distances.
It can be noticed that the differences
range from a few picoseconds at higher
perihelion values ($r_0=10 \ R_{\odot}$)
to roughly 400 ps at solar superior
conjunctions. Moreover, for
a given perihelion or closest approach,
these picosecond deviations remain
practically the same over the solar
system distances. Conversely, these discrepancies are
significantly reduced in Figure \ref{fig2}(b)
where our exact equation (\ref{Shaptimed}) is
compared to the second-order STD equation (\ref{Shap2ndGR}).
As a result, the picosecond-level differences between
both expressions are considerably reduced.

\begin{figure}
\begin{center}
\includegraphics[width=.70\textwidth]{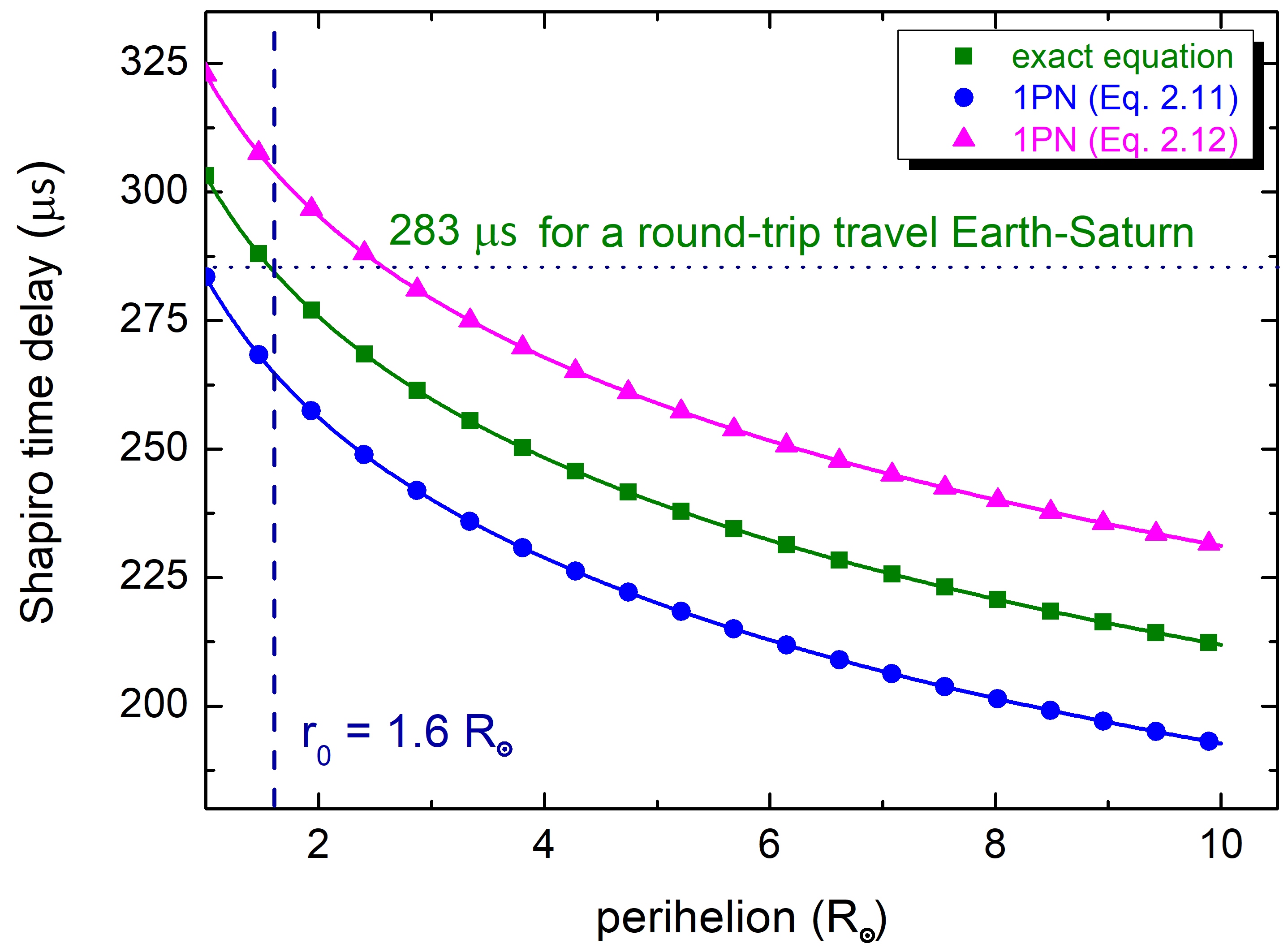}
\caption{Shapiro time delay versus
the perihelion $r_0$ for a light beam describing
a round-trip travel from the Earth to
Saturn, where the STD has been calculated
via our exact equation (\ref{Shaptimed})
in isotropic coordinates, Misner's 1PN
equation (\ref{Shap1PNisoMis}), and Malta's 1PN
equation (\ref{Shap1PNisoMal}). It can be noticed
microsecond differences in the STD for perihelions
up to $10 \ R_{\odot}$, with a maximum deviation of
39.31 $\mu\textrm{s}$ between both 1PN expressions
for $r_0=1 \ R_{\odot}$.}\label{fig3}
\end{center}
\end{figure}

Furthermore, let us now compare our theoretical
results with the 1PN equations for the STD, where a suitable
coordinate transformation is accomplished, according to equations
(\ref{transr0}), (\ref{transrP}) and (\ref{transrE}).
Thus, in Figure \ref{fig3},
we show the Shapiro time delay versus
the perihelion $r_0$ for a light beam describing
a round-trip travel from Earth to
Saturn, where the STD has been calculated
via our exact equation (\ref{Shaptimed})
in isotropic coordinates, Misner's 1PN
equation (\ref{Shap1PNisoMis}), and Malta's 1PN
equation (\ref{Shap1PNisoMal}). In this case,
microsecond discrepancies are found
for perihelions up to $10 \ R_{\odot}$ with
a maximum deviation of 39.31 $\mu\textrm{s}$
between both 1PN expressions for $r_0=1 \ R_{\odot}$.
However, as reported by Bertotti \emph{et al.} \cite{BE03} in his
famous time delay measurements where the Saturn orbiter
Cassini was used as a reflector, the minimum
allowed perihelion was $1.6 \ R_{\odot}$ with
an estimated STD of 288 $\mu\textrm{s}$
\cite{ZS22,ZS24}. It can be observed that our
exact analytical results (i.e., computed via equation (\ref{Shaptimed})
in isotropic coordinates) fit the experimental data
fairly well. In addition, it should be addressed
that our 1PN results are in accordance with the previously commented
apparent discordances between both 1PN analytical expressions.

\begin{figure}
\begin{center}
\includegraphics[width=1.0\textwidth]{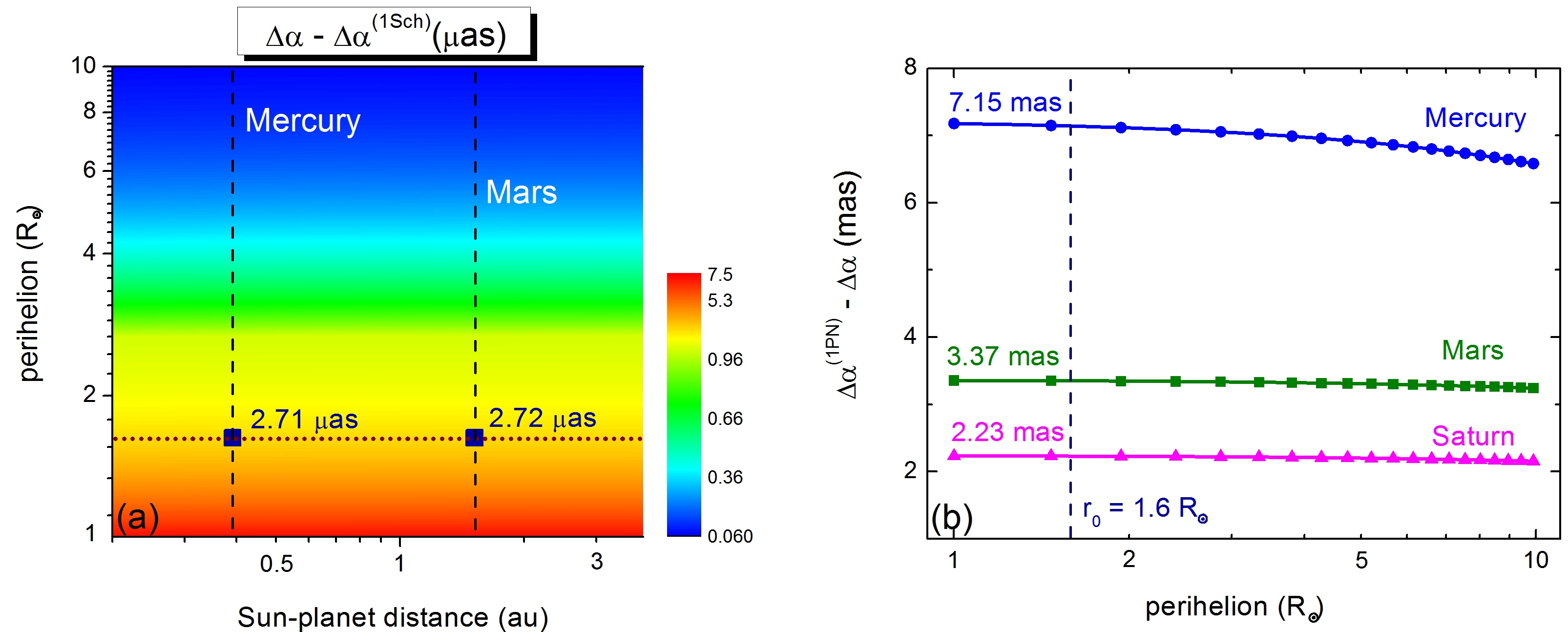}
\caption{(a) Absolute difference between our
exact equation (\ref{GDGOSch}) and the
first-order GDA in Schwarzschild
coordinates, equation (\ref{GDSch1st}),
as a function of the perihelion and
the Sun-planet distance and (b) absolute
difference between our analytical
equation (\ref{GDGOiso}) and the 1PN formula,
equation (\ref{GD1PN}), both in isotropic
coordinates.}\label{fig4}
\end{center}
\end{figure}

For completeness, let us now analyze the
gravitational deflection angle of light in
the context of planetary solar conjunctions
(please, see again Figure \ref{fig1}). In this
way, the absolute difference between
our exact equation (\ref{GDGOSch}) and
the first-order GDA in Schwarzschild
coordinates, equation (\ref{GDSch1st}),
as a function of the perihelion and
the Sun-planet distance, is depicted
in Figure \ref{fig4}(a). It is noteworthy that
the GDA discrepancies lie below the
microarcosecond ($\mu\textrm{as}$) level, which is
much lower than the accuracy of current
observatories such as GAIA \cite{BR21}.
Nevertheless, when this GDA comparison is
performed between our exact
equation (\ref{GDGOiso}) and the 1PN formula,
equation (\ref{GD1PN}), both in isotropic
coordinates, these differences increase
notably. As shown in Figure \ref{fig4}(b),
milliarcosecond (mas) deviations are achieved
for different planets in the solar system,
which is consistent with previous reported GDA
calculations \cite{BA24}.

\section{Discussion}

In summary, an exact equation for the
Shapiro time delay in a Schwarzschild
spacetime was derived, providing
accurate values of the STD
in comparison with previous first and second-order
expressions, where picosecond and sub-picosecond
accuracy levels are desired.
In relation to these STD approximate formulas,
it should be mentioned the consistent asymptotic
behaviour of the second-order term in (\ref{Shap2ndGR})
when $r_{\rm P}\to\infty$, unlike the first-order equation (\ref{Shap1stGR})
where a divergent logarithmic term is obtained when planet P
is placed at an infinite distance from the Sun.

It is also worth noting that our STD calculations
are restricted to perihelion values
below $10 \ R_{\odot}$, provided the astrophysical
situation illustrated in Figure \ref{fig1}, where
the planets are near their superior solar
conjunction. Nonetheless, the alleged STD
discrepancies between our exact equation
and the former first-order expressions remain
unchanged for higher perihelion distances.
Even in the case that our massive body
is a giant planet like Jupiter, the lowest
STD discrepancies between the monopolar
approximate equation (\ref{Shap1stGR})
and our exact equation (\ref{Shaptimed}),
for a light beam describing a round-trip
travel from the Earth to Saturn,
is about 1.17 ps at $r_0=1.6 \ R_{\rm J}$,
where $R_{\rm J}$ denotes Jovian radius.

A plausible explanation for the largest
STD deviations reported in this article
is that, as stated by Ballmer \emph{et al.} \cite{BA10},
the gravitational time delay may be different
in alternative theories of gravity than
Einstein's general relativity. In this way,
our exact equation (\ref{Shaptimed}) for
the STD is derived directly from
the Schwarzschild metric, whereas the 1PN approximate
expressions come from the post-Newtonian
approximation, which alternatively expresses
Einstein's (nonlinear) equations of gravity in
terms of the lowest-order deviations from
Newton's law of universal gravitation.
Moreover, non-negligible discrepancies in
the propagation time delay between analytical
equations (in Schwarzschild coordinates)
and first-order post-Newtonian expressions
(in harmonic coordinates) have been
reported in the context of pulsars orbiting
a supermassive black hole \cite{HA19},
where an appropriate coordinate system
transformation was performed.

We must bear in mind that our exact formulas
for the Shapiro time delay and gravitational
deflection angle of light can only be
applied to astrophysical situations
where the central massive body is spherical
and static. Obviously, this represents an
unrealistic physical scenario for the solar
system, so more complex gravity theories
(such as the post-Newtonian and
post-post-Newtonian formalisms)
are needed to include higher-order effects
such as the non-sphericity, slow motion, or
spin-multipoles of the gravitational masses
\cite{ZS22,ZS24}. Nevertheless, our exact
equations (expressed in isotropic coordinates)
might help reduce the 1PN deviations presented here.

In conclusion, our analytical expressions
for the Shapiro time delay should improve
the time delay calculations in ongoing and
future research, particularly in the solar system
or extragalactic astrophysical scenarios, where a
considerable accuracy level is desirable.

\ack{The author would like to acknowledge
Konstantinos Glampedakis for helpful
comments and assistance with manuscript
preparation, and Sven Zschocke for interesting
discussions on post-Newtonian formalism. O. del Barco
also thanks research support from Agencia Estatal
de Investigaci\'{o}n (PID2019-105684RB-I00,
PID2020-113919RBI00).}

\section*{Data availability statement}

All data and numerical results generated or analyzed
during this study are included in this article
(and based on the references therein). The
analytical calculations performed in the Appendices
were carried out using Mathematica
software and are fully available from the
corresponding author upon reasonable request.

\section*{Conflicts of interest}

The authors declare no conflict of interest.

\section*{ORCID iDs}

Oscar del Barco https://orcid.org/0000-0001-7502-9164 \\

\appendix

\section{Propagation time calculation in terms of elliptic integrals}

The exact analytical expression for the coordinate
time $t$, in Schwarzschild coordinates, can be analytically
obtained by the direct integration of equation (\ref{DEGRdef})
via a Wolfram Mathematica software. For this purpose, it is
convenient to perform some changes of variable which will
facilitate the analytical computation.

First, let us introduce the parameter $\lambda=r_{\rm s}/r_0$
into equation (\ref{DEGRdef})
\begin{equation}\label{cdtror1}
c\mathrm{d}t=\frac{\mathrm{d}r \ \sqrt{1-\lambda}}{\left(1-\frac{r_{\rm s}}{r}\right)
\sqrt{1-\left(\frac{r_0}{r}\right)^{2}+\lambda\left[\left(\frac{r_0}{r}\right)^{3}-1\right]}},
\end{equation}
and then carry out the final change $r_0/r=x$, so the differential
equation (\ref{cdtror1}) can be rewritten as
\begin{equation}\label{cdtrordef}
c\mathrm{d}t=\frac{\mathrm{d}x \ (-r_0 \sqrt{1-\lambda})}{(x^2-\lambda x^3)
\sqrt{1-x^2+\lambda(x^3-1)}} =\mathrm{d}x \ (-r_0 \sqrt{1-\lambda}) \ A(x,\lambda).
\end{equation}
After an appropriate integration process, the primitive function
resulted to be
\begin{eqnarray}\label{primgen}
c t(x)&=&A_0(x)+A_{1}(x) \ F(z(x),k) +
A_{2}(x) \ E(z(x),k)
            \nonumber\\
&+& A_{31}(x) \ \Pi(m_1;z(x),k)+ A_{32}(x) \ \Pi(m_2;z(x),k),
\end{eqnarray}
where $Q^2=1+2\lambda-3\lambda^2$ and $F(z(x),k)$, $E(z(x),k)$ and $\Pi(m_{i};z(x),k)$
represent the Legendre elliptic integrals of the first,
second, and third kinds, respectively, with the
following relations for the Jacobi amplitude $z(x)$
\begin{equation}\label{zrr0}
z(x) = \iu \ \mathrm{arcsinh}\sqrt{\frac{2\lambda(x-1)}{-1+3\lambda+Q}},
\end{equation}
elliptic modulus $k$
\begin{equation}\label{kr0}
k=\frac{1-3\lambda-Q}{1-3\lambda+Q},
\end{equation}
and the two additional parameters for the third
kind elliptical integrals
\begin{equation}\label{m1}
m_1=\frac{-1+3\lambda+Q}{2(-1+\lambda)},
\end{equation}
\begin{equation}\label{m2}
m_2=\frac{-1+3\lambda+Q}{2\lambda}.
\end{equation}
Furthermore, the extra factors included
in equation (\ref{primgen}) are expressed as
\begin{equation}\label{A0x}
A_0(x)=r_0 \sqrt{\frac{1-x^2+(-1+x^3)\lambda}{x^2(1-\lambda)}},
\end{equation}

\begin{equation}\label{A1x}
A_{1}(x)=\frac{r_0(-1+\lambda-Q)}{4\sqrt{1-\lambda}} \ S(x),
\end{equation}

\begin{equation}\label{A2x}
A_{2}(x)=\frac{r_0(1-3\lambda+Q)}{4\sqrt{1-\lambda}} \ S(x),
\end{equation}

\begin{equation}\label{A31x}
A_{31}(x)=\frac{-r_0\lambda^2}{\sqrt{1-\lambda}} \ S(x),
\end{equation}

\begin{equation}\label{A32x}
A_{32}(x)=-r_0 \lambda \sqrt{1-\lambda} \ S(x),
\end{equation}
where
\begin{equation}\label{Sr0}
S(x)= \sqrt{\frac{2(-1+\lambda+2x\lambda)^2-2Q^2}
{\lambda(1-3\lambda+Q)(-1+\lambda+x(-1+\lambda+x\lambda))}}.
\end{equation}

To derive the exact equation for
the propagation time of a light beam
from its perihelion $r_0$ to a given
position $r$, $(\Delta t)_{r_0 \to r}$,
the following relation should be considered
\begin{equation}\label{coordtimer0rgen}
(\Delta t)_{r_0 \to r}= t(r)- t(r_0)=t(x)- t(x=1),
\end{equation}
therefore, equation (\ref{primgen}) must be
evaluated at $x=1$. After some simple algebra, one observes
that $t(x=1)=0$, so equation (\ref{coordtimer0rgen})
is finally given by
\begin{eqnarray}\label{coordtimer0rdeff}
(\Delta t)_{r_0 \to r}&=& \frac{1}{c} \left[A_0(x)+A_{1}(x) \ F(z(x),k) +
A_{2}(x) \ E(z(x),k) \right.
            \nonumber\\
&+& \left.A_{31}(x) \ \Pi(m_1;z(x),k)+ A_{32}(x) \ \Pi(m_2;z(x),k)\right],
\end{eqnarray}
where the Schwarschild coordinate $r$ is related to the dimensionless
factor $x$ via the expression $x=r_0/r$.

\section{Second-order expression for the STD in a Schwarzschild spacetime}

Let us now expand the RHS terms of equation (\ref{cdtrordef})
up to second order in $\lambda$
\begin{eqnarray}\label{Axlambdaexp}
A(x,\lambda)&=&\frac{1}{x^2\sqrt{1-x^2}}+\left(\frac{1+3x(1+x)}
{2x^2 \sqrt{1-x}(1+x)^{3/2}}\right) \ \lambda
\nonumber\\
&+& \left(\frac{3+5x(1+x)(2+3x(1+x))}{8x^2
\sqrt{1-x}(1+x)^{5/2}}\right) \ \lambda^2 + \mathrm{O}[\lambda]^{3},
\end{eqnarray}
and
\begin{equation}\label{sqrtexp}
\sqrt{1-\lambda}=1-\frac{\lambda}{2}-\frac{\lambda^2}{8}
+ \mathrm{O}[\lambda]^{3}.
\end{equation}

Then, we integrate the different terms of $A(x,\lambda)$ and
multiply each primitive by the expanded external factor $\sqrt{1-\lambda}$.
Neglecting higher-order terms, the final primitive function has the form
\begin{eqnarray}\label{ctrsecond}
c t(x)&=&\frac{-r_0}{8x(1+x)^{3/2}}\left[\sqrt{1-x}(-8(1+x)^{2}-4x(1+x)
\lambda+x(4+5x)\lambda^2)\right.
\nonumber\\
&+& \left.x(1+x)^{3/2}\left(8\lambda\log\left(\frac{x}{1+\sqrt{1-x^2}}\right)+
15\lambda^2\arctan\left(\frac{x}{\sqrt{1-x^2}}\right)\right)\right],
\nonumber\\
\end{eqnarray}
where, again, $x=r_0/r$. Evaluating equation (\ref{ctrsecond}) at $x=1$
(that is, at the perihelion $r_0$) we have
\begin{equation}\label{ctrperi2nd}
c t(x=1)=\frac{-15\pi r_0 \lambda^2}{16},
\end{equation}
so the propagation time from $r_0$ to a given
position $r$ up to second order in $\lambda$ can be written as
(please, see equation (\ref{coordtimer0rgen}))
\begin{eqnarray}\label{ctrseconddef}
(\Delta t)_{r_0 \to r}&=&\frac{-r_0}{8c x(1+x)^{3/2}}\left[\sqrt{1-x}(-8(1+x)^{2}-
4x(1+x)\lambda+x(4+5x)\lambda^2)\right.
\nonumber\\
&+& \left.x(1+x)^{3/2}\left(8\lambda\log\left(\frac{x}{1+\sqrt{1-x^2}}\right)
+15\lambda^2\arctan\left(\frac{x}{\sqrt{1-x^2}}\right)\right)\right]
\nonumber\\
&+& \frac{15\pi r_0 \lambda^2}{16c}.
\end{eqnarray}

Accordingly, the second-order STD expression in Schwarschild coordinates
for a light pulse propagating from $r_{\rm P}$ to $r_{\rm E}$ is given by
\begin{eqnarray}\label{Delta2Scha}
\Delta t^{(\rm 2Sch)}&=& 2\left[(\Delta t)_{r_0 \to r_{\rm P}} + (\Delta t)_{r_0 \to r_{\rm E}}
-\frac{1}{c} \left(\sqrt{r_{\rm P}^2-r_0^2}+\sqrt{r_{\rm E}^2-r_0^2}\right)\right]
\nonumber\\
&=& \frac{r_0\left(\sqrt{1-x_{\rm P}}(4x_{\rm P}(1+x_{\rm P})
\lambda-x_{\rm P}(4+5x_{\rm P})\lambda^2)\right)}{4c x_{\rm P}(1+x_{\rm P})^{3/2}}
\nonumber\\
&-& \frac{r_0}{4c} \left[8\lambda\log\left(\frac{x_{\rm P}}{1+\sqrt{1-x_{\rm P}^2}}\right)
+15\lambda^2\arctan\left(\frac{x_{\rm P}}{\sqrt{1-x_{\rm P}^2}}\right)\right]
\nonumber\\
&+& \frac{r_0\left(\sqrt{1-x_{\rm E}}(4x_{\rm E}(1+x_{\rm E})
\lambda-x_{\rm E}(4+5x_{\rm E})\lambda^2)\right)}{4c x_{\rm E}(1+x_{\rm E})^{3/2}}
\nonumber\\
&-& \frac{r_0}{4c} \left[8\lambda\log\left(\frac{x_{\rm E}}{1+\sqrt{1-x_{\rm E}^2}}\right)
+15\lambda^2\arctan\left(\frac{x_{\rm E}}{\sqrt{1-x_{\rm E}^2}}\right)\right]
\nonumber\\
&+& \frac{15\pi r_0 \lambda^2}{4c},
\end{eqnarray}
which can be rewritten in the simplified form
\begin{eqnarray}\label{Delta2Schfin}
\Delta t^{(\rm 2Sch)}&=&\frac{2r_{\rm s}}{c}
\left[\log\left(\frac{r_{\rm P}+\sqrt{r_{\rm P}^{2}-r_0^2}}{r_0}\right)+
\log\left(\frac{r_{\rm E}+\sqrt{r_{\rm E}^{2}-r_0^2}}{r_0}\right)\right]
\nonumber\\
&+& \frac{r_{\rm s}}{c} \left(\sqrt{\frac{r_{\rm P}-r_0}{r_{\rm P}+r_0}}
+\sqrt{\frac{r_{\rm E}-r_0}{r_{\rm E}+r_0}}\right)+\frac{r_{\rm s}^{2}}{4cr_0}
\left[15\arctan\left(\frac{\sqrt{r_{\rm P}^{2}-r_0^{2}}}{r_0}\right)\right.
\nonumber\\
&-& \left.\left(\frac{4r_{\rm P}+5r_0}{r_{\rm P}+r_0}\right)
\sqrt{\frac{r_{\rm P}-r_0}{r_{\rm P}+r_0}}+
15\arctan\left(\frac{\sqrt{r_{\rm E}^{2}-r_0^{2}}}{r_0}\right)\right.
\nonumber\\
&-& \left.\left(\frac{4r_{\rm E}+5r_0}{r_{\rm E}+r_0}\right)
\sqrt{\frac{r_{\rm E}-r_0}{r_{\rm E}+r_0}}\right].
\end{eqnarray}

\bigbreak


\section*{References}

\begin{thebibliography}{90}

\bibitem{SH64} Shapiro I I 1964 Fourth Test of General Relativity
{\it Phys. Rev. Lett.} \textbf{13} 789

\bibitem{SH68} Shapiro I I, Pettengill G H, Ash M E, Stone M L,
Smith W B, Ingalls R P and Brockelman R A 1968 Fourth Test of General
Relativity: Preliminary Results {\it Phys. Rev. Lett.} \textbf{20} 1265

\bibitem{SH71} Shapiro I I, Ash M E, Ingalls R P, Smith W B, Campbell D B, Dyce
R B, Jurgens R F and Pettengill G H 1971 Fourth test of general relativity:
new radar result {\it Phys. Rev. Lett.} \textbf{26} 1132

\bibitem{KO01} Kopeikin S M 2001 Testing the relativistic effect of the propagation
of gravity by very long baseline interferometry {\it Astrophys. J. Lett.} \textbf{556} L1-L5

\bibitem{KO03} Kopeikin S M 2003 The post-Newtonian treatment of the VLBI experiment
on September 8, 2002 {\it Phys. Lett. A} {\bf 312} 147

\bibitem{WI14} Will C M 2014 The Confrontation between General Relativity and
Experiment {\it Living Rev. Relativ.} {\bf 17} 4

\bibitem{AS08} Asada H 2008 Gravitational time delays along multiple light
paths as a probe of physics beyond Einstein Gravity {\it Phys. Lett. B}
{\bf 661} 78

\bibitem{ED21} Edelstein J D, Ghosh R, Laddha A and Sarkar S 2021
Causality constraints in Quadratic Gravity {\it J. High Energy Phys.}
{\bf 2021} 150

\bibitem{DY22} Dyadina P I and Labazova S P 2022 On Shapiro time
delay in massive scalar-tensor theories {\it J. Cosmol. Astropart. Phys.}
{\bf 2022} 029

\bibitem{HA19} Hackmann E and Dhani A 2019 The propagation delay in the
timing of a pulsar orbiting a supermassive black hole {\it Gen. Relativ. Gravit.}
{\bf 51} 37

\bibitem{CR20} Cromartie H T \emph{et al.} 2020 Relativistic Shapiro delay
measurements of an extremely massive millisecond pulsar {\it Nat. Astron.}
{\bf 4} 72

\bibitem{BE22} Ben-Salem B and Hackmann E 2022 Propagation time delay and
frame dragging effects of lightlike geodesics in the timing of a pulsar
orbiting SgrA* {\it Mon. Not. R. Astron. Soc} {\bf 516} 1768–1780

\bibitem{BA58} Balazs N L 1958 Effect of a gravitational field, due to a
rotating body, on the plane of polarization of an electromagnetic
wave {\it Phys. Rev.} {\bf 110} 236

\bibitem{PL60} Plebanski J 1960 Electromagnetic waves in gravitational fields
{\it Phys. Rev.} {\bf 118} 1396

\bibitem{DE71} De Felice F 1971 On the gravitational field acting as an
optical medium {\it Gen. Relativ. Gravit.} {\bf 2} 347

\bibitem{NA95} Nandi K K and Islam A 1995 On the optical-mechanical
analogy in general relativity {\it Am. J. Phys.} {\bf 63} 251

\bibitem{EV96a} Evans J, Nandi K K and Islam A 1996 The optical-mechanical
analogy in general relativity: Exact Newtonian forms for the equations of
motion of particles and photons {\it Gen. Relativ. Gravit.} {\bf 28} 413

\bibitem{EV96b} Evans J, Nandi K K and Islam A 1996 The optical–mechanical
analogy in general relativity: New methods for the paths of light and
of the planets {\it Am. J. Phys.} {\bf 64} 1404

\bibitem{SE10} Sen A K 2010 A more exact expression for the gravitational
deflection of light, derived using material medium approach
{\it Astrophysics} {\bf 53} 560

\bibitem{RO15} Roy S and Sen A K 2015 Trajectory of a light ray in
Kerr field: a material medium approach {\it Astrophys. Space Sci.}
{\bf 360} 23

\bibitem{RO17} Roy S and Sen A K 2017 Deflection of light ray due
to a charged body using material medium approach {\it Z. Naturforsch. A}
{\bf 72} 1113

\bibitem{FE19} Feng G and Huang J 2019 A geometric optics method for
calculating light propagation in gravitational fields
{\it Optik} {\bf 194} 163082

\bibitem{FE20} Feng G and Huang J 2020 An optical perspective on the
theory of relativity- II: Gravitational deflection of light and
Shapiro time delay {\it Optik} {\bf 224} 165685

\bibitem{BA24} Barco O 2024 An accurate equation for the gravitational
bending of light by a static massive object
{\it Mon. Not. R. Astron. Soc} {\bf 535} 2504–2510

\bibitem{RU25} Ruggiero M L 2025 Efects of gravitational waves on
electromagnetic fields {\it Eur. Phys. J. C} {\bf 85} 1

\bibitem{RO25} Roy S, Kala S, Singha A, Nandan H and Sen A K 2025
Deflection of light due to Kerr Sen black hole in heterotic string
theory using material medium approach {\it Eur. Phys. J. C} {\bf 85} 772

\bibitem{ZS22} Zschocke S 2022 Time delay in the quadrupole field
of a body at rest in 2PN approximation {\it Phys. Rev. D} {\bf 106} 104052

\bibitem{ZS24} Zschocke S 2024 Time delay in the gravitational field
of an axisymmetric body at rest with full mass and spin multipole
structure {\it Phys. Rev. D} {\bf 109} 064044

\bibitem{BA10} Ballmer S, M\'{a}rka S and Shawhan P 2010 Feasibility
of measuring the Shapiro time delay over meter-scale distances
{\it Class. Quantum Grav.} {\bf 27} 185018

\bibitem{SU20} Sullivan A G, Veske D, M\'{a}rka Z, Bartos I, Ballmer S,
Shawhan P and M\'{a}rka S 2020 Can we use next-generation gravitational
wave detectors for terrestrial precision measurements of Shapiro delay?
{\it Class. Quantum Grav.} {\bf 37} 205005

\bibitem{WE72} Weinberg S 1972 {\it Gravitation and cosmology: principles and
applications of the general theory of relativity} (John Wiley and Sons, New York)

\bibitem{WA84} Wald R M 1984 {\it General Relativity}
(The University of Chicago Press, Chicago)

\bibitem{MA24} Malta P C and Zarro C A D 2024 Bounds on the photon
mass via the Shapiro effect in the solar system
{\it Phys. Rev. D} {\bf 110} 095017

\bibitem{MI73} Misner C W, Thorne K S and Wheeler J A 1973 {\it Gravitation}
(Princeton University Press, San Francisco)

\bibitem{PO14} Poisson E and Will C M 2014 {\it Gravity: Newtonian, Post-Newtonian, Relativistic}
(Cambridge University Press, Cambridge)

\bibitem{LI22a} Li Y \emph{et al.} 2022 Light deflection under the gravitational
field of Jupiter—testing general relativity {\it Astrophys. J.} {\bf 925} 47

\bibitem{LI22b} Li Y, Xu Y, Bian S, Lin Z, Li J, Liu D and Chaojie H 2022
The Effect of Light Deflection by Solar System Objects on High-precision Square
Kilometre Array Astrometry {\it Astrophys. J.} {\bf 938} 58

\bibitem{BE03} Bertotti B, Iess L and Tortora P 2003 A test of general
relativity using radio links with the Cassini spacecraft
{\it Nature} {\bf 425} 374

\bibitem{BR21} Brown A 2021 Microarcsecond Astrometry: Science Highlights from Gaia
{\it Annu. Rev. Astron. Astrophys} {\bf 59} 59

\end{thebibliography}
\end{document}